\documentclass{emulateapj}
\usepackage{apjfonts}

\usepackage{amsmath}

\usepackage{graphics}

\newcommand{\bd}{\begin{displaymath}}
\newcommand{\ed}{\end{displaymath}}

\slugcomment{accepted by ApJ}

\shorttitle{Growth of massive black holes during radiatively inefficient accretion phases}

\shortauthors{Cao}

\begin{document}

\title{Growth of massive black holes during radiatively inefficient accretion phases}

\author{Xinwu Cao}
\affil{Shanghai Astronomical Observatory, Chinese Academy of
Sciences, 80 Nandan Road, Shanghai, 200030, China; Email:
xwcao@shao.ac.cn}

\begin{abstract}

We derive the black hole mass density as a function of redshift from the hard X-ray AGN 
luminosity function assuming that massive black holes grow via accreting the circumnuclear 
gases. The derived black hole mass density matches the measured local black hole mass density 
at $z=0$, which requires the average radiative efficiency of AGNs to be $\sim 0.1-0.17$. The 
massive black holes in most faint active galactic nuclei (AGNs) and even normal galaxies are 
still accreting gases, though their accretion rates are very low. Radiatively inefficient 
accretion flows (RIAFs) are supposed in these faint sources, which should radiate mostly in 
the hard X-ray band. We calculate the contribution to the X-ray background from both the bright 
AGNs and the RIAFs in faint AGNs/normal galaxies. Our calculations show that both the observed 
intensity and spectral shape of the XRB with an energy peak at $\sim 30$~keV can be well 
reproduced without including the emission of Compton-thick AGNs, if the massive black holes in 
faint AGNs/normal galaxies are spinning rapidly with $a\sim 0.9$ and accreting at rates 
$\dot{m}\sim 1.0-3.0\times 10^{-4}$.  It indicates that less than $\sim$5 per cent of local 
massive black hole mass density was accreted during radiatively inefficient accretion phases, 
which is obviously only an upper limit, because Compton-thick AGNs have not been considered. 
If the same number of the Compton-thick AGNs with $\log N_{\rm H}=24-25$ as those with 
$\log N_{\rm H}=23-24$ is considered, the fraction of local black hole mass density accumulated 
during inefficient accretion phases should be lower than $\sim$2 per cent. The constraints 
of the XRB can provide upper limits on the average accretion rate for inactive galaxies.

\end{abstract}


\section{Introduction}

There is evidence that most nearby galaxies contain massive black holes at
their centers, and a tight correlation was revealed between central massive black hole mass
and the velocity dispersion of the galaxy \citep[][]{fm00,g00}, which strongly
suggests co-evolution of massive black holes and their host galaxies. This correlation,
together with the correlation between black hole mass and the host galactic bulge
luminosity, are widely used to estimate the black hole masses in galaxies.

The growth of massive black holes may probably be linked to accretion processes \citep{s82}.
\citet{yt02} estimated the black hole masses from the stellar velocity dispersions of
galaxies measured by the Sloan Digital Sky Survey (SDSS) using the empirical relation
between black hole mass and the velocity dispersion, and the local black hole mass density
was derived. They further calculated the black hole mass density accreted during optical bright
quasar phases using an optical quasar luminosity function (LF), and found that the accreted mass density
is consistent with the local black hole mass density estimated from the velocity dispersions, if 
a radiative efficiency $\sim 0.1$ is adopted for quasars. 
This implies that the growth of massive black holes may be dominantly through accretion during
optically bright quasar phases. The optical quasar LF is directly linked to the accretion 
history of massive black holes, however,
it has overlooked many faint AGNs and type II AGNs. The 
X-ray luminosity functions (XLFs) derived from the surveys in the soft X-ray band ($\la$
3~keV) may have missed many obscured (type II) AGNs, while the
hard X-ray surveys ($\sim$ 2-10~keV) can trace the whole AGN
population including obscured type II AGNs and extend to low X-ray
luminosity \citep{u03}, thus providing a useful tool to explore the
accretion history of AGNs in the universe. Many works on XLFs were
carried out in either the soft X-ray band ($\la 3$~keV)
\citep*[e.g.,][]{m91,b93,p97,mhs00} or the hard X-ray band ($\ga
2$~keV) \citep*[e.g.,][]{b98,c03,u03}. 

The cosmological X-ray background (XRB) is mostly contributed by AGNs \citep{h98,s98}. 
In the most popular synthesis models of the XRB based on the
unification schemes for AGNs, the cosmological XRB from $\la
2$~keV to more than several hundred keV can be fairly well
reproduced by using a series of template spectra of AGNs with different 
obscuration, which are derived by assuming an intrinsic AGN spectrum 
consisting of a power-law X-ray spectrum with an exponential cutoff around several
hundred keV \citep*[see, e.g.,][]{mf94,mgf94,c95,grs99,u03}.
\citet{df97} have alternatively proposed that the hard XRB above 10 keV may be dominated by
the thermal bremsstrahlung emission from the advection dominated
accretion flows (ADAFs) in low-luminosity AGNs. Further detailed
ADAF spectral calculations \citep{d99} have shown that many sources at
redshift $z\sim 2-3$ with ADAFs accreting at the rates close to
the critical value would be required to reproduce the observed XRB spectral shape 
with an energy peak at $\sim 30$~keV.  

Recently, \citet{u03} derived a hard X-ray
luminosity function (HXLF) from a highly complete AGN sample
(2-10~keV), which includes both type I and type II AGNs (except
Compton-thick AGNs). Based on this HXLF, their synthesis models
can explain most of the observed XRB from the soft X-ray band to
the hard X-ray band around several hundred keV. Their calculations
slightly ($\approx$10--20\%) underestimate the relative shape of
the XRB spectrum around its peak intensity (see Fig. 18 in their
work). Such a discrepancy can be explained provided the same
number of Compton-thick AGNs with $\log N_{\rm H}=24-25$ as those
with $\log N_{\rm H}=23-24$ is included. 
As the number density of Compton-thick AGNs is still quite uncertain, 
the residual XRB could be attributed to both the Compton-thick AGNs with 
$\log N_{\rm H}=24-25$ and the RIAFs in faint AGNs/normal galaxies. Although their 
relative importance is still unclear, it 
is obvious that the spectral shape of the observed XRB cannot be solely 
attributed to the RIAFs/ADAFs in Seyferts.

There are a variety of studies exploring the evolution of AGNs
based on either optical quasar LFs or XLFs, or both
\citep*[e.g.,][]{hr93,hm00,kh00,yt02,wl03,ma04,h05c}. A common
conclusion is that the timescale of AGN activities is short
compared with the Hubble timescale, though the quantitative
results on the bright quasar lifetime vary from $\sim10^{7}$ to
$\sim10^{9}$ years  for different investigations.
The standard optically thick accretion disks are present in bright
AGNs, as their accretion rates are high. The AGN activity may be
switched off while the gases near the black hole are exhausted
\citep*[see][for a review, and the references therein]{n02}.
When the accretion rate $\dot{m}$ ($\dot{m}=\dot{M}/\dot{M}_{\rm
Edd}$) declines below a critical value $\dot{m}_{\rm crit}\sim 0.01$, the
standard disk transits to a RIAF \citep*[e.g.,][]{ny95}. RIAFs
are optically thin, very hot, and their spectra are peaked in the
hard X-ray band. In a previous work, \citet{cao05} calculated the contribution
to the hard XRB from the RIAFs in the faint AGNs. It was found that
the accretion rate $\dot{m}$ must decrease rapidly to far below the critical
rate $\dot{m}_{\rm crit}$ after the accretion mode transition, otherwise,
the RIAFs may produce too much hard X-ray emission to match the observed hard
XRB. This implies that the growth of massive black holes during
radiatively inefficient accretion phases is not important. \citet{hnh06}
considered the distribution of local supermassive black hole Eddington ratios
and accretion rates, and they found that black hole mass growth, both of the
integrated mass density and the masses of most individual objects, are
dominated by an earlier, radiatively efficient, high accretion rate stage, and
not by the radiatively inefficient low accretion rate phases.

In this paper, we will explore quantitatively on how much mass 
in the local massive black holes was accreted during their radiatively inefficient accretion phases 
from the constraint of the XRB. The cosmological parameters $\Omega_{\rm M}=0.3$,
$\Omega_{\Lambda}=0.7$, and $H_0=70~ {\rm km~s^{-1}~Mpc^{-1}}$
have been adopted in this work.

\section{Black hole mass density}

The HXLF given by \citet{u03} is so far most suitable for our
present investigation, as it includes both type I and type II AGNs
(except Compton-thick AGNs).  The HXLF in 2$-$10 keV derived by \citet{u03} is
\begin{equation}
 \frac{{\rm d} \Phi
(L_{\rm X}, z)}{{\rm d Log} L_{\rm X}} = \frac{{\rm d} \Phi
(L_{\rm X}, 0)}{{\rm d Log} L_{\rm X}} e(z, L_{\rm X}),
\label{hxlf} \end{equation}
where \begin{equation} \frac{{\rm d} \Phi (L_{\rm X},
z=0)}{{\rm d Log} L_{\rm X}} = A [(L_{\rm X}/L_{*})^{\gamma 1} +
(L_{\rm X}/L_{*})^{\gamma 2}]^{-1}, \end{equation}
 \begin{equation}
 e(z, L_{\rm X}) =
\begin{cases}
(1+z)^{p1} & (z<z_{\rm c}(L_{\rm X})) \\
e(z_{\rm c})[(1+z)/(1+z_{\rm c}(L_{\rm X}))]^{p2} & (z \geq z_{\rm
c}(L_{\rm X})),
\end{cases}
\end{equation} and \begin{equation} z_{\rm c}(L_{\rm X}) =
\begin{cases}
 z_{\rm c}^* & (L_{\rm X} \geq L_a) \\
z_{\rm c}^* (L_{\rm X}/L_a)^\alpha & (L_{\rm X}<L_a).
\end{cases}
\end{equation}
All the parameters for our present adopted cosmology are as
follows: $A=5.04(\pm0.33)\times10^{-6}~{\rm Mpc^{-3}}$, $\log
L_{*} ({\rm ergs~s^{-1}})=43.94^{+0.21}_{-0.26}$,
$\gamma_1=0.86\pm0.15$, $\gamma_2=2.23\pm0.13$, $p1=4.23\pm0.39$;
$p2=-1.5$, $z_{\rm c}=1.9$, and $\log L_{a} ({\rm
ergs~s^{-1}})=44.6$ \citep{u03}. All the sources described by \citet{u03}'s HXLF
have hard X-ray luminosities $L_{\rm X}^{2-10~{\rm keV}}$ 
between $10^{41.5}-10^{48}$ ergs~s$^{-1}$. In this work, we refer to the sources
described by the \citet{u03}'s HXLF as active galaxies, while all the remainder
with $L_{\rm X}^{2-10~{\rm keV}}<10^{41.5}$~ergs~s$^{-1}$ as inactive galaxies.

The cosmological evolution of black hole mass density caused by accretion during 
active galaxy phases is described by 
\begin{equation}
{\frac {{\rm d}\rho_{\rm bh}^{\rm A}(z)}{{\rm d}z} }=
{\frac {{\rm d}t}{{\rm d}z}}{1\over {\rm M_\odot}}
\int\limits_{41.5}^{48}
{\frac {(1-\epsilon)f_{\rm cor}L_{\rm X}}{\epsilon c^2}}
\frac{{\rm d} \Phi (L_{\rm X},
z)}{{\rm d Log} L_{\rm X}} {\rm d Log} L_{\rm X},~~~  \label{rhobha}
\end{equation}
where $\Phi (L_{\rm X}, z)$ is the HXLF given by Eq. (\ref{hxlf}), $\rho_{\rm bh}^{\rm A}(z)$ 
(in units of ${\rm M_\odot~Mpc^{-3}}$) is 
the black hole mass density accreted during active galaxy phases between $z$ and $z_{\rm max}$, and 
$\epsilon$ is the radiative efficiency for AGNs. The ratio $f_{\rm cor}$ of 
the bolometric luminosity to the X-ray luminosity in $2-10$~keV is given by   
\begin{equation}
\log f_{\rm cor}=1.54+0.24{\cal L}+0.012{\cal L}^2-0.0015{\cal L}^3, \label{fcor}
\end{equation}
where ${\cal L} = \log L_{\rm bol}/L_\odot-12$, and $L_{\rm bol}$ is the bolometric 
luminosity \citep{ma04}.

The central massive black holes in inactive galaxies may still 
be accreting gases, though the accretion rates may be very low as their
hard X-ray luminosities $L_{\rm X}^{2-10~{\rm keV}}$ are below $10^{41.5}$ ergs~s$^{-1}$. 
The Eddington accretion rate 
$\dot{M}_{\rm Edd}=(M_{\rm bh}/{\rm M_\odot})L_{\odot,\rm Edd}/\eta_{\rm eff}c^2$, 
where $L_{\odot,\rm Edd}=1.38\times10^{38} {\rm ergs~s^{-1}}$, and a conventional radiative 
efficiency $\eta_{\rm eff}=0.1$ is adopted here. This radiative efficiency $\eta_{\rm eff}$ 
is only for the 
definition of the Eddington rate $\dot{M}_{\rm Edd}$, and it can be different from 
the real radiative efficiency $\epsilon$ in Eq. (\ref{rhobha}). 
When the mass accretion rate
$\dot{m}$ ($\dot{m}=\dot{M}/\dot{M}_{\rm
Edd}$) is below a critical value $\dot{m}_{\rm crit}$, the
standard disk converts to a RIAF \citep*[e.g.,][]{ny95}. RIAFs should be present in most, if not all,  of those inactive galaxies.
Besides the black hole mass density $\rho_{\rm bh}^{\rm A}(z)$ accreted in active galaxies, 
we need to calculate the black hole density $\rho_{\rm bh}^{\rm B}(z)$ accreted in inactive 
galaxies. Assuming an average dimensionless mass accretion rate $\dot{m}_{\rm inact}^{\rm aver}$ for
those inactive galaxies, we can calculate the black hole mass density
$\rho_{\rm bh}^{\rm B}(z)$ accreted during  
inactive galaxy phases between $z$ and $z_{\rm max}$ by
\begin{equation}
{\frac {{\rm d} \rho_{\rm bh}^{\rm B}(z)}{{\rm d}z}}
={\frac   {\rho_{\rm bh}^{\rm inact}(z)\dot{m}_{\rm inact}^{\rm aver}L_{\odot,\rm Edd}
(1-\epsilon^{\rm RIAF})} {{\rm M}_\odot\eta_{\rm eff}c^2}} {\frac {{\rm d}t}{{\rm d}z}}, ~~~  \label{rhobhb}
\end{equation}
where $\epsilon^{\rm RIAF}$ is the radiative efficiency of the RIAFs in inactive galaxies, 
provided the black hole mass density of inactive
galaxies $\rho_{\rm bh}^{\rm inact}(z)$ (in units of ${\rm M_\odot~Mpc^{-3}}$) as a function of 
redshift $z$ is known. Here, $\rho_{\rm bh}^{\rm inact}(z)$ is the black hole mass density 
of inactive galaxies at $z$, while $\rho_{\rm bh}^{\rm B}(z)$ describes the mass density   
accreted during inactive galaxy phases between $z$ and $z_{\rm max}$.  
The radiative efficiency $\epsilon^{\rm RIAF}$ is usually much lower than 
that for standard thin disks. Our numerical calculations for RIAF spectra show that $\epsilon^{\rm RIAF}$ is significantly lower than 
0.01, so we approximate $\epsilon^{\rm RIAF}\simeq 0$ in Eq. (\ref{rhobhb}) while calculating the black hole mass 
density.  

The total black hole mass density $\rho_{\rm bh}^{\rm acc}(z)$ accumulated through 
accretion both in active and inactive galaxies is: 
\begin{equation}
\rho_{\rm bh}^{\rm acc}(z)=\rho_{\rm bh}^{\rm A}(z)+\rho_{\rm bh}^{\rm B}(z), \label{rhobhtot}
\end{equation}
i.e., a sum of black hole mass densities accreted in active and inactive galaxy phases. 
In order to calculate $\rho_{\rm bh}^{\rm acc}(z)$, we need to know
the black hole mass density of inactive galaxies $\rho_{\rm bh}^{\rm inact}(z)$ as 
a function of $z$ (see Eq. \ref{rhobhb}).
Assuming the growth of massive black holes is dominated by accretion \citep[e.g.,][]{yt02}, 
the total black hole mass density is given by 
\begin{equation}
\rho_{\rm bh}(z)\simeq
\rho_{\rm bh}^{\rm acc}(z)+\rho_{\rm bh}(z_{\rm max})
=\rho_{\rm bh}^{\rm A}(z)+\rho_{\rm bh}^{\rm B}(z)+\rho_{\rm bh}(z_{\rm max}), \label{rhobhtotb}
\end{equation} 
where $\rho_{\rm bh}(z_{\rm max})$ 
is the total black hole mass density at $z_{\rm max}$. The black hole mass density for
active galaxies in the co-moving space at redshift $z$ can be calculated from the XLF by
\begin{equation}
\rho^{\rm act}_{\rm bh}(z)={\frac {1}{(\epsilon/\eta_{\rm eff})\dot{m}^{\rm aver}_{\rm act}L_{\rm Edd,\odot}}}
\int\limits^{48}_{41.5}f_{\rm cor}L_{\rm X}\frac{{\rm d} \Phi (L_{\rm X},
z)}{{\rm d Log} L_{\rm X}} {\rm d Log} L_{\rm X}~~~ {\rm
M}_{\odot} {\rm Mpc^{-3}}, \label{bhmdensact}
\end{equation}
where $\Phi (L_{\rm X}, z)$ is the HXLF given by Eq. (\ref{hxlf}),
$\dot{m}^{\rm aver}_{\rm act}$ is the average dimensionless accretion rate
for the active galaxies. The black hole mass density 
for inactive galaxies $\rho_{\rm bh}^{\rm inact}(z)$ can be derived by subtracting that 
for active galaxies $\rho_{\rm bh}^{\rm act}(z)$ from the total black hole density 
$\rho_{\rm bh}(z)$. The black hole
mass density of inactive galaxies $\rho_{\rm bh}^{\rm inact}(z)$
can therefore be calculated by
\begin{equation}
\rho_{\rm bh}^{\rm inact}(z)=\rho_{\rm bh}(z)-\rho_{\rm bh}^{\rm act}(z)
=\rho_{\rm bh}^{\rm A}(z)+\rho_{\rm bh}^{\rm B}(z)+\rho_{\rm bh}(z_{\rm max})-\rho_{\rm bh}^{\rm act}(z). 
\label{rhobhinact}
\end{equation}
Assuming all galaxies are active at $z=z_{\rm max}$ \citep*[e.g.,][]{ma04}, the black hole mass 
density $\rho_{\rm bh}(z_{\rm max})=\rho_{\rm bh}^{\rm act}(z_{\rm max})$ is available 
by using Eq. (\ref{bhmdensact}). At $z=z_{\rm max}$, 
$\rho_{\rm bh}^{\rm A}=\rho_{\rm bh}^{\rm B}=0$. Thus, we integrate Eqs. (\ref{rhobha}) and  
(\ref{rhobhb}) simultaneously 
over $z$ from $z_{\rm max}$ using Eqs. (\ref{bhmdensact}) and (\ref{rhobhinact}), and the black hole mass 
densities as functions of $z$ are finally available provided the three parameters: $\dot{m}^{\rm aver}_{\rm act}$, 
$\dot{m}^{\rm aver}_{\rm inact}$, and $\epsilon$ are specified. We adopt 
$z_{\rm max}=5$ and $\dot{m}^{\rm aver}_{\rm act}=0.1$ in all our calculations. 
The value of the radiative efficiency $\epsilon$ is tuned to let the derived 
total black hole mass density $\rho_{\rm bh}(z)$  match 
$\rho_{\rm bh}^{\rm local}$ at $z=0$, for a given $\dot{m}_{\rm inact}^{\rm aver}$.

The local black hole mass density $\rho_{\rm bh}^{\rm local}$ was estimated from the stellar velocity dispersion measured 
with the SDSS by using the relation between black hole mass and the stellar velocity 
dispersion. The resulting local black hole mass densities vary from $\sim 2\times 10^5$ to $5\times 10^5 {\rm M_\odot~Mpc^{-3}}$
\citep*[e.g.,][]{s99,yt02,s03,ma04}. In some previous works, the black hole mass densities as 
functions of 
redshift were derived to match the measured local black hole mass density 
by integrating AGN LFs \citep*[e.g.,][]{yt02,ma04}.  
The difference between these previous works and ours is that 
we have included the contribution to the local black hole mass density from accretion in 
inactive galaxies.

Our calculations are carried out 
for two different values of the local black hole mass density:  
$\rho_{\rm bh}^{\rm local}=2.5\times 10^5 {\rm M_\odot~Mpc^{-3}}$, and 
$4.0\times 10^5 {\rm M_\odot~Mpc^{-3}}$, respectively, considering the different values of  
$\rho_{\rm bh}^{\rm local}$ derived in the literature \citep*[e.g.,][]{s99,yt02,s03,ma04}. 
We find that the radiative efficiencies  
$\epsilon\sim 0.07-0.12$ are required, so that the total black hole mass densities accumulated 
through accretion can match the measured local black hole mass densities (see Figs. \ref{fig1}, and Table 1 for a summary 
of the results). The derived radiative efficiency depends only weakly on 
$\dot{m}_{\rm inact}^{\rm aver}$, because most of black hole mass was 
accreted during active galaxy phases. 
If the average mass accretion rate $\dot{m}_{\rm inact}^{\rm aver}$ is 
high, a significant fraction of local black hole mass density was accumulated 
in inactive galaxies ($\sim$16 per cent, for $\dot{m}_{\rm inact}^{\rm aver}=0.001$, see 
Table 1).   
If $\dot{m}_{\rm inact}^{\rm aver}\la 3\times 10^{-4}$, less than $\sim$ 5 per cent of 
local black hole mass density was accreted in inactive galaxies.  
At low redshifts with $z\la 2$, the black hole mass density is dominated by 
the black holes in inactive galaxies (see Fig. \ref{fig2}). 
In Fig. \ref{fig2}. we find that the derived  
black hole mass densities $\rho_{\rm bh}^{\rm inact}(z)$ in inactive galaxies at low redshifts  
mainly depend on the value of the measured local black hole mass 
density $\rho_{\rm bh}^{\rm local}$ adopted.  
\figurenum{1}
\centerline{\includegraphics[angle=0,width=7.8cm]{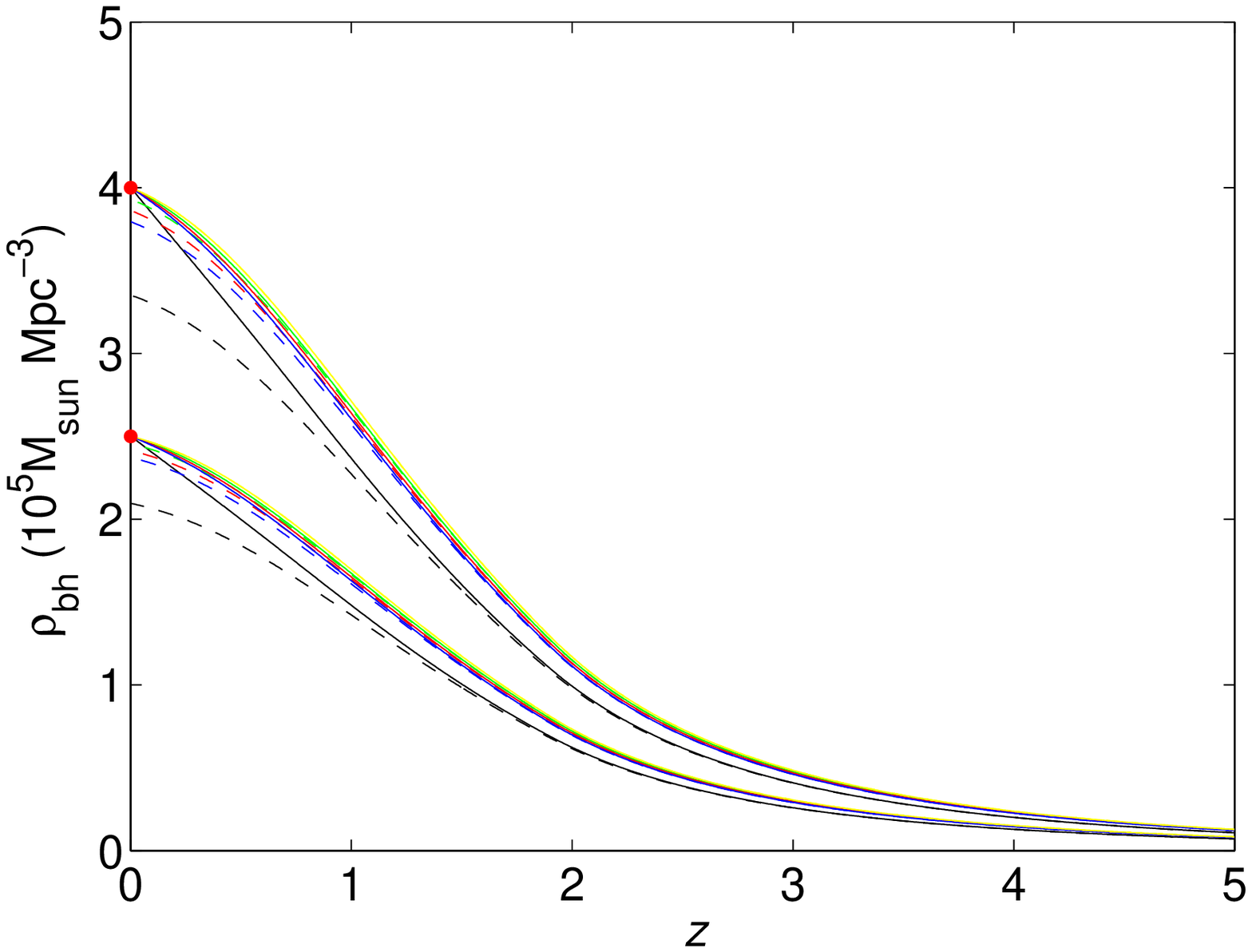}}
\figcaption{The total black hole mass densities $\rho_{\rm bh}(z)$
accumulated through accretion as functions of redshift $z$ (solid lines), while the black hole 
mass densities accreted during active galaxy phases only [$\rho_{\rm bh}^{\rm A}(z)
+\rho_{\rm bh}(z_{\rm max})$]
are plotted as dashed lines. 
The upper set of curves is derived for $\rho_{\rm bh}^{\rm local}=4.0\times 10^{5} {\rm M_\odot~Mpc^{-3}}$, 
while the lower one is for $\rho_{\rm bh}^{\rm local}=2.5\times 10^{5} {\rm M_\odot~Mpc^{-3}}$. 
The different colors correspond to different models:
black (Model A), blue (Model B), red (Model C), green (Model D), and yellow (Model E). The description of
different models is summarized in Table 1. \label{fig1}}  \centerline{}
\figurenum{2}
\centerline{\includegraphics[angle=0,width=7.8cm]{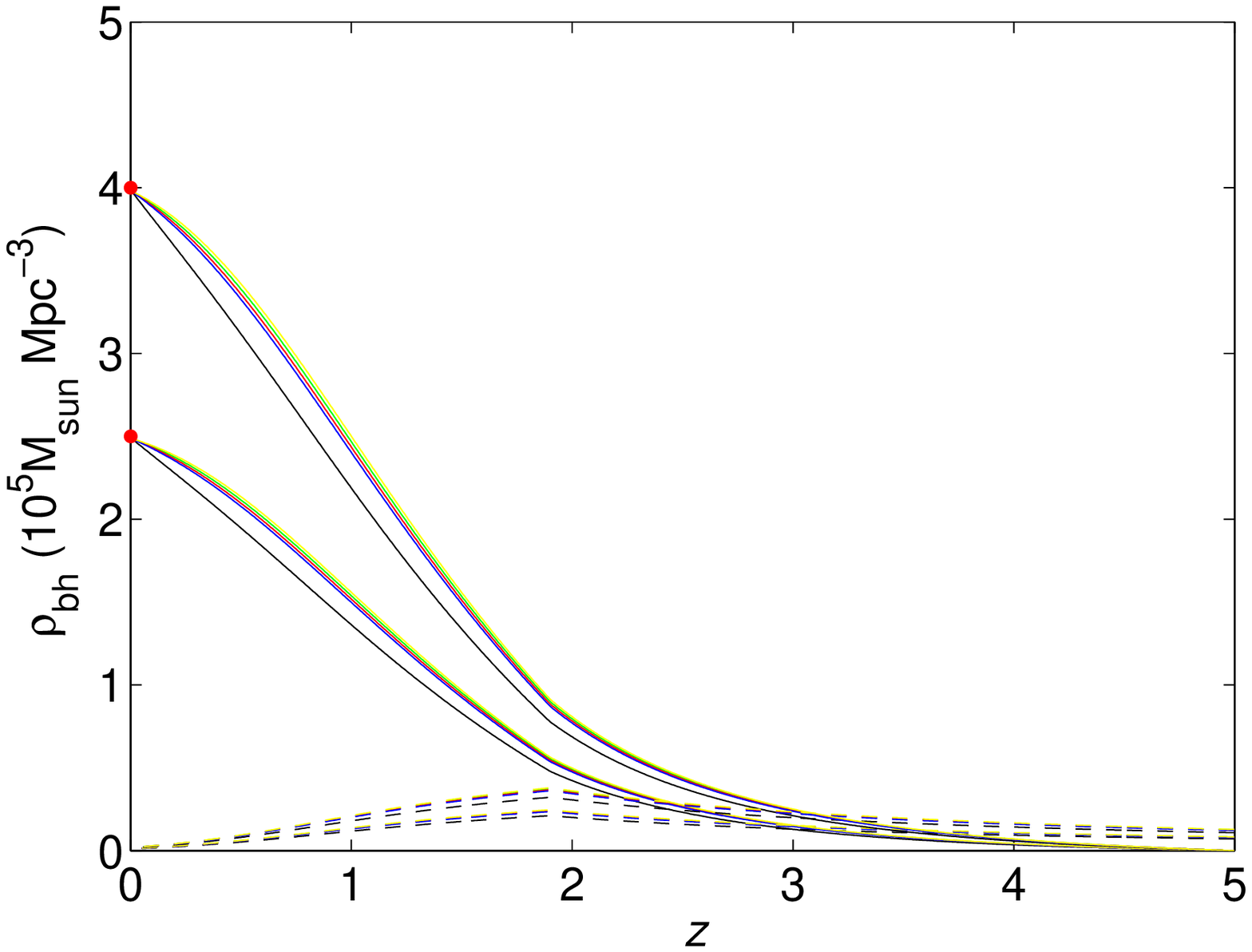}}
\figcaption{The black hole mass densities 
$\rho_{\rm bh}^{\rm inact}(z)$ for inactive galaxies 
as functions of redshift $z$ (solid lines), while the dashed lines represent the black hole 
mass densities $\rho_{\rm bh}^{\rm act}(z)$ in active galaxies. The upper set of curves is derived for $\rho_{\rm bh}^{\rm local}=4.0\times 10^{5} {\rm M_\odot~Mpc^{-3}}$, 
while the lower one is for $\rho_{\rm bh}^{\rm local}=2.5\times 10^{5} {\rm M_\odot~Mpc^{-3}}$. 
The different colors correspond to different models:
black (Model A), blue (Model B), red (Model C), green (Model D), and yellow (Model E). The description of
different models is summarized in Table 1.
\label{fig2}}\centerline{}

\section{RIAF spectra}

The RIAF spectrum is almost proportional to black hole mass $M_{\rm bh}$ for a given 
dimensionless mass accretion rate $\dot{m}$. 
In order to calculate the contribution of the RIAFs in all inactive galaxies to the
XRB, we need to use the X-ray spectrum $l_{\rm X}(E)$
of a RIAF around a $10^8{\rm M}_\odot$ black hole accreting at the rate 
$\dot{m}_{\rm inact}^{\rm aver}$ as a template spectrum. We employ the approach
suggested by \citet{m00} to calculate the global structure of an
accretion flow surrounding a massive black hole in the general relativistic frame,
which allows us to calculate the structure of an accretion flow surrounding either a spinning or
non-spinning black hole. All the radiation processes
are included in the calculations of the global accretion flow
structure \citep*[see][for details and the references
therein]{m00}. However, the values of some parameters adopted are different 
from those in \citet{m00} \citep[see discussion in \S 5, and][]{cao05}. 
In the spectral calculations, the gravitational
redshift effect is considered, while the relativistic optics near
the black hole is neglected. This will not affect our final
results on the XRB, as the inactive galaxies should have randomly
distributed orientations and the stacked spectra would not be
affected by the relativistic optics.

The global structure of a RIAF surrounding a black hole spinning at $a$ with 
mass $M_{\rm bh}$ can be calculated, if some parameters, $\dot{m}$, $\alpha$, 
$\beta$ and $\delta$, are specified. The parameter $\delta$ describes 
the fraction of the viscously dissipated energy directly going into 
electrons in the accretion flow.  
The three-dimensional MHD simulations suggest that the viscosity
parameter $\alpha$ in the accretion flows is $\sim 0.1$
\citep{a98}, or $\sim 0.05-0.2$ \citep{hb02}.
The critical accretion rate
$\dot{m}_{\rm crit}\simeq 0.01$ is suggested by different authors either from
observations or theoretical arguements 
\citep*[see][for a review, and the references therein]{n02}. Our numerical calculations for the global structure
of the flows show $\dot{m}_{\rm crit}\simeq 0.01$ for $\alpha=0.2$. Thus, we adopt
$\alpha=0.2$ in this work. The parameter $\beta$, defined as 
$p_{\rm m}=B^2/8\pi=(1-\beta)p_{\rm tot}$ ($p_{\rm tot}=p_{\rm gas}+p_{\rm m}$),  
describes the magnetic field strength of the accretion flow. This parameter is in fact not a free
parameter, which is related to the viscosity parameter $\alpha$ as
$\beta\simeq (6\alpha-3)/(4\alpha-3)$, as suggested by the MHD simulations \citep{hgb96,nmq98}.
For $\alpha=0.2$, $\beta\simeq 0.8$ is required. It was pointed out that a
significant fraction of the viscously dissipated energy could go into electrons by magnetic
reconnection, if the magnetic fields in the flow are strong
\citep{bl97,bl00}. In our calculations, we adopt a conventional value of 
$\delta=0.1$ in all the calculations. We plot the X-ray spectra of the RIAFs 
accreting at different rates for Schwarzschild holes in Fig. \ref{fig5}. The spectra 
of the accretion flows surrounding the Kerr black holes with $a=0.9$ are ploted in  
Fig. \ref{fig6}. 
\figurenum{3}
\centerline{\includegraphics[angle=0,width=7.8cm]{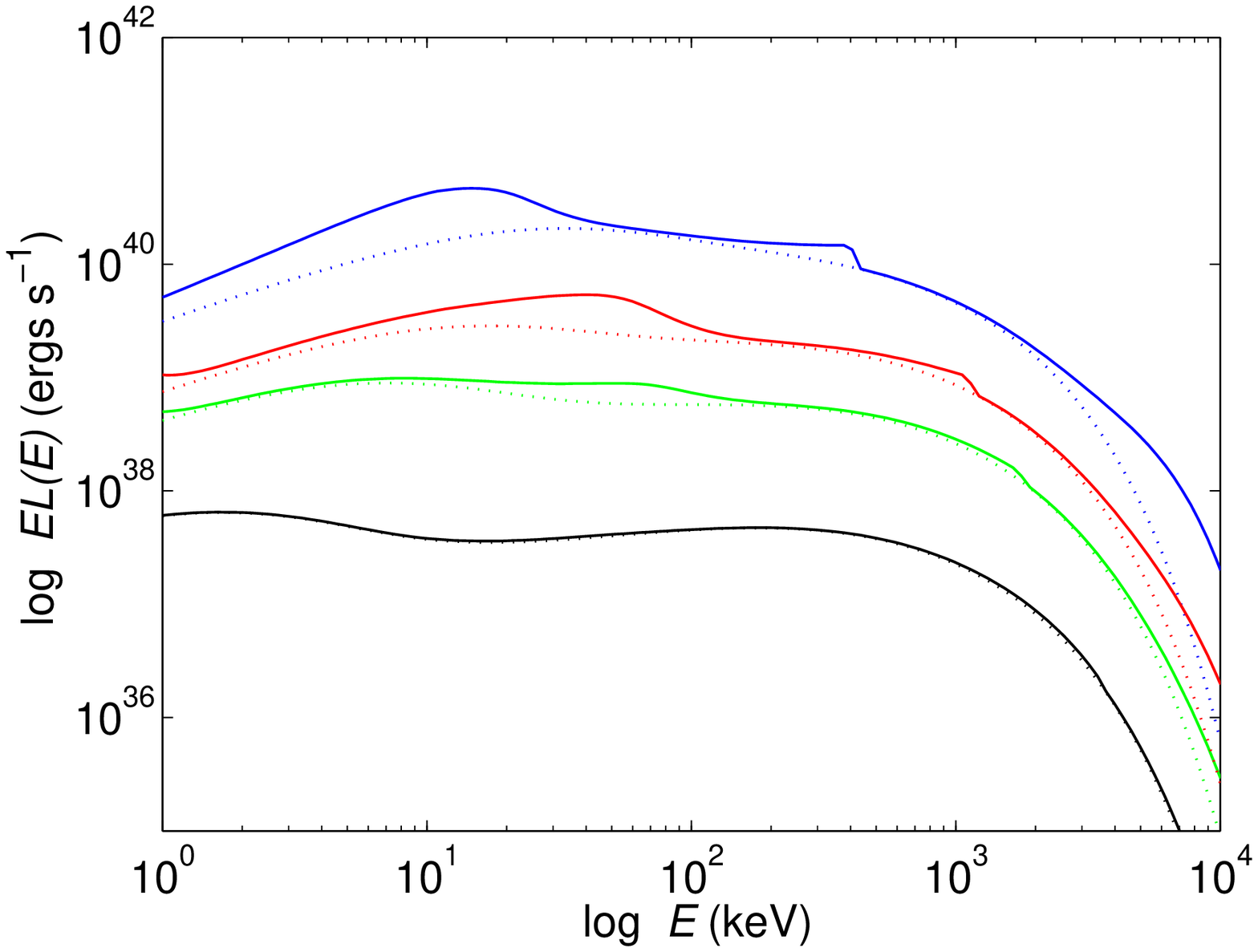}}
\figcaption{The spectra of the RIAFs surrounding Schwarzschild black holes with 
$10^8 {\rm M}_\odot$  accreting at different rates: $\dot{m}=1.0\times 10^{-3}$ (blue lines), $5.0\times 10^{-4}$ (red lines),
$3.0\times 10^{-4}$ (green lines), and $1.0\times 10^{-4}$ (black lines).
The viscosity parameter $\alpha=0.2$, the fraction of the
magnetic pressure $1-\beta=0.2$, and the fraction of viscously dissipated energy directly 
heating the electrons $\delta=0.1$, are adopted in the calculations. The solid lines 
represent the spectra of the RIAFs, while the dotted
lines represent their bremsstrahlung spectra only.
\label{fig5}}\centerline{}
\figurenum{4}
\centerline{\includegraphics[angle=0,width=7.8cm]{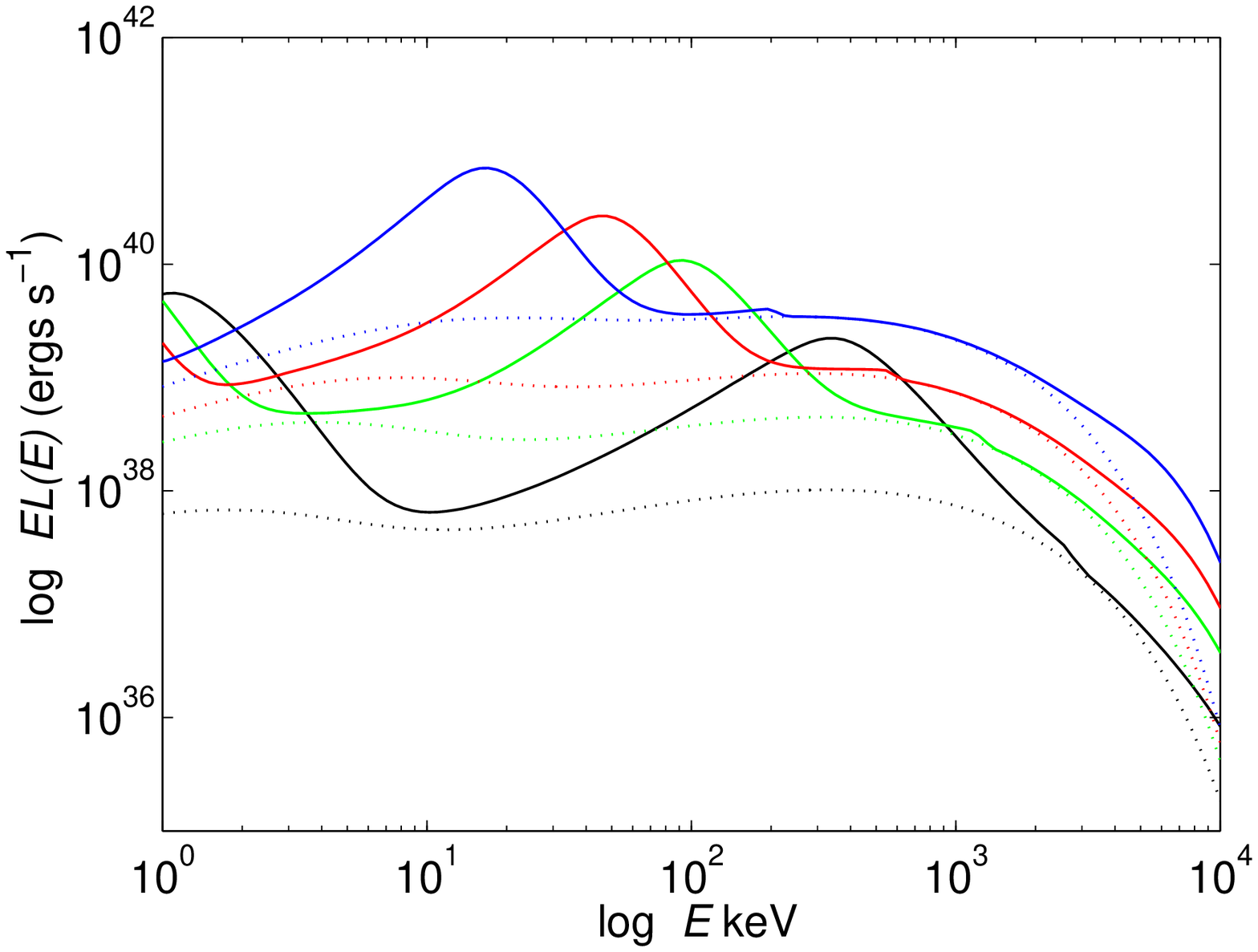}}
\figcaption{The spectra of the RIAFs surrounding Kerr black holes ($a=0.9$) 
with $10^8 {\rm M}_\odot$ accreting at different rates: $\dot{m}=5.0\times 10^{-4}$ (blue lines), 
$3.0\times 10^{-4}$ (red lines),
$2.0\times 10^{-4}$ (green lines), and $1.0\times 10^{-4}$ (black lines).
The same disk parameters are adopted in the calculations as those in Fig. \ref{fig5}. The solid lines 
represent the spectra of the RIAFs, while the dotted
lines represent their bremsstrahlung spectra only. 
\label{fig6}}\centerline{}

\section{Contribution to the XRB from the RIAFs in inactive galaxies}

The contribution of the RIAFs in all inactive galaxies to the
cosmological XRB can be calculated by
\begin{equation} f_{\rm X}(E)
={\frac 1{10^{8}{\rm M_\odot} } }
\int\limits_{0}^{z_{\rm max}}
{\frac {\rho_{\rm bh}^{\rm inact}(z)(1+z)l_{\rm X}[(1+z)E]} {4\pi
d_{\rm L}^2}}{\frac {{\rm d}V}{{\rm d}z}}{\rm d}z,
\label{xrb}
\end{equation}
where $l_{\rm X}(E)$ is the spectrum of a RIAF surrounding a $10^8 {\rm M_\odot}$
black hole accreting at $\dot{m}_{\rm inact}^{\rm aver}$, which is calculated as described
in \S 3.

There is no doubt that the contribution from active galaxies
is important to the XRB, as the {\it BeppoSAX} observations
showed that the power-law X-ray spectra of bright AGNs extends to
several hundred keV \citep[e.g.,][]{n00}. 
\citet{u03} adopted a template intrinsic AGN spectrum, i.e., a power law$+$an exponential spectrum 
with cutoff energy $E_{\rm c}$ of 500 keV, in their XRB synthesis model calculations. 
They also included the contribution to the XRB by the Compton-thick AGNs with 
$\log N_{\rm H}=24-25$ assuming a fixed ratio to those with $\log N_{\rm H}=23-24$ without 
considering the redshift dependence of the ratio.  
The effects of Compton scattering leading to 
significant decrease of the emitted flux of the nucleus even in the hard X-ray bands have been 
properly considered in their calculations, while the contribution of the sources 
with $\log N_{\rm H}>25$ is neglected by assuming that all X-rays are absorbed. 
In this work, we simply adopt the results of their XRB synthesis model calculations 
for AGNs. We consider that
the XRB consists not only of the emission from type I/II bright
AGNs (Compton-thin) described by the HXLF, but also of the
emission from RIAFs in those inactive galaxies that are not included in
this HXLF. All curves of the XRB synthesis for Compton thin/thick AGNs ploted in the 
figures of this work are taken from \citet{u03}. 
In Figs. \ref{fig9} and \ref{fig10}, we plot the XRBs contributed by both 
active galaxies and inactive galaxies with different parameters adopted, and 
compare them with the observed XRB. Here, the contribution from Compton-thick AGNs has not been 
included, which may be unrealistic, because Compton-thick AGNs have already been 
detected, though their number density is still quite uncertain. However, these 
calculations, at least, provide strict constraints (upper limits) on the average mass accretion 
rates of inactive galaxies. We define a parameter $\xi$ to describe the number ratio of 
Compton-thick AGNs with $\log N_{\rm H}=24-25$ to those with $\log N_{\rm H}=23-24$. 
In Fig. \ref{fig12}, we plot the results 
of the XRB synthesis, in which the contribution of Compton-thick AGNs is considered by assuming 
$\xi=0.5$ and 1.0, respectively. For a given $\xi$, the average accretion rate 
$\dot{m}_{\rm inact}^{\rm aver}$ is tuned in order not to overproduce the observed 
XRB in 10-30 keV \citep{u03}, Thus, the constraints (upper limits) 
on the average accretion rates $\dot{m}_{\rm inact}^{\rm aver}$ for inactive galaxies from the cosmological 
XRB are plotted as functions of $\xi$ in Fig. \ref{fig13}. 
\figurenum{5}
\centerline{\includegraphics[angle=0,width=7.8cm]{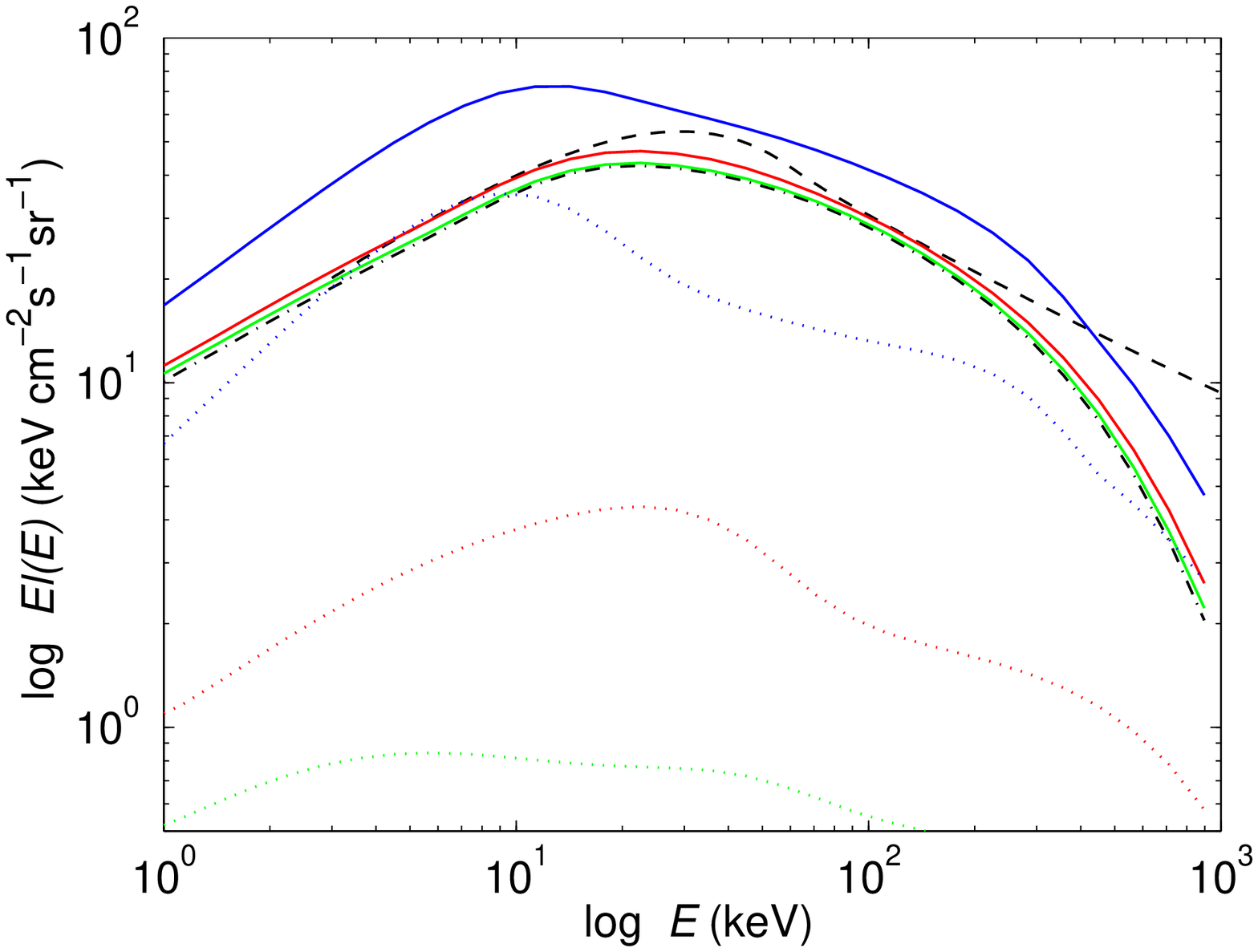}}
\figcaption{The contribution to the XRB from active galaxies
and the RIAFs surrounding Schwarzschild black holes in inactive galaxies.
The dashed line is the observed XRB \citep[taken from][]{u03}. The dot-dashed 
line represents the contribution by type I/II AGNs (Compton-thin), which is taken from
\citet{u03}. The lines with different colors represent the XRB synthesis model calculations for different values 
of the average accretion rates for inactive galaxies: 
$\dot{m}_{\rm inact}^{\rm aver}=1.0\times 10^{-3}$ (blue),
$5.0\times 10^{-4}$ (red), and $3.0\times 10^{-4}$ (green).
The solid lines represent the XRB contributed by type I/II (Compton-thin) active galaxies decribed by the HXLF and the RIAFs in all inactive
galaxies, while the dotted lines are only for the contributions from the RIAFs in all 
inactive galaxies. The black hole mass density for inactive galaxies 
$\rho_{\rm bh}^{\rm inact}(z)$ used in the XRB synthesis 
calcualtions is derived for $\rho_{\rm bh}^{\rm local}=4.0\times 10^{5} {\rm M_\odot~Mpc^{-3}}$.
\label{fig9}}\centerline{}
\figurenum{6}
\centerline{\includegraphics[angle=0,width=7.8cm]{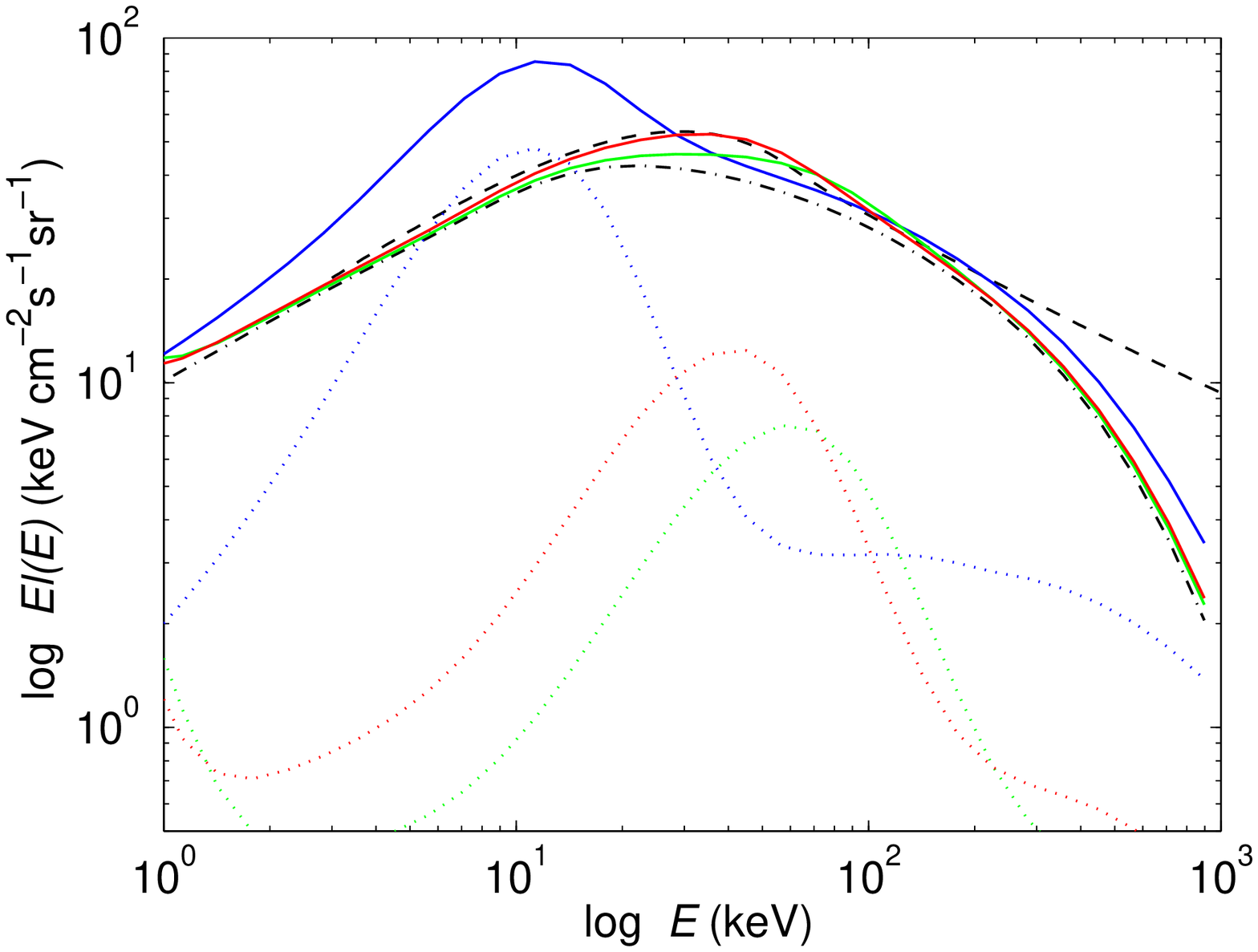}}
\figcaption{The same as Fig. \ref{fig9}, but for the RIAFs surrounding
the Kerr black holes with $a=0.9$ in inactive galaxies. The lines
with different colors represent different values of the average accretion
rates: $\dot{m}_{\rm inact}^{\rm aver}=5.0\times 10^{-4}$ (blue),
$2.5\times 10^{-4}$ (red), and $2.0\times 10^{-4}$ (green).
\label{fig10}}\centerline{}
\figurenum{7}
\centerline{\includegraphics[angle=0,width=7.8cm]{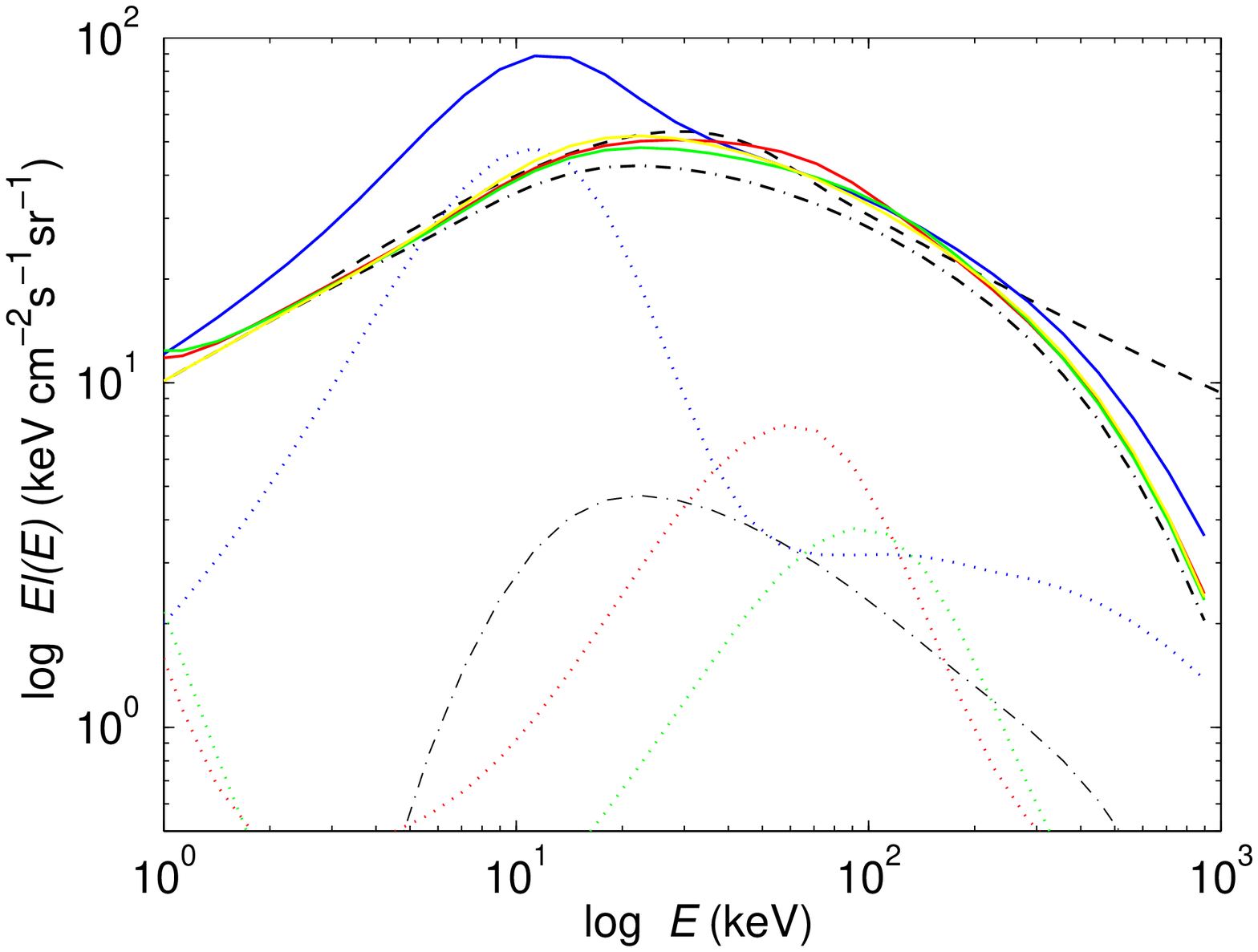}}
\figcaption{The same as Fig. \ref{fig9}, but for the Kerr black holes 
with $a=0.9$ and the contribution of Compton-thick AGNs is included. The thin dash-dotted 
line represents the contribution of Compton-thick AGNs \citep[taken from][]{u03}. The yellow line 
represents the synthesis model for type I/II$+$Compton-thick AGNs assuming $\xi=1.0$, 
in which the contribution from the RIAFs in inactive galaxies is not included 
\citep[taken from][]{u03}.  All lines 
with other colors correspond to the XRB synthesis model calculations by assuming $\xi=0.5$, 
i.e., the number of Compton-thick AGNs with $\log N_{\rm H}=24-25$ is half of those with 
$\log N_{\rm H}=23-24$, These lines represent different values of the average accretion rates: 
$\dot{m}_{\rm inact}^{\rm aver}=5.0\times 10^{-4}$ (blue),
$2.0\times 10^{-4}$ (red), and $1.5\times 10^{-4}$ (green). The local black hole mass 
density $\rho_{\rm bh}^{\rm local}=4.0\times 10^{5} {\rm M_\odot~Mpc^{-3}}$ is adopted to 
derive the black hole mass density $\rho_{\rm bh}^{\rm inact}(z)$ in inactive galaxies. 
\label{fig12}}\centerline{}
\figurenum{8}
\centerline{\includegraphics[angle=0,width=7.8cm]{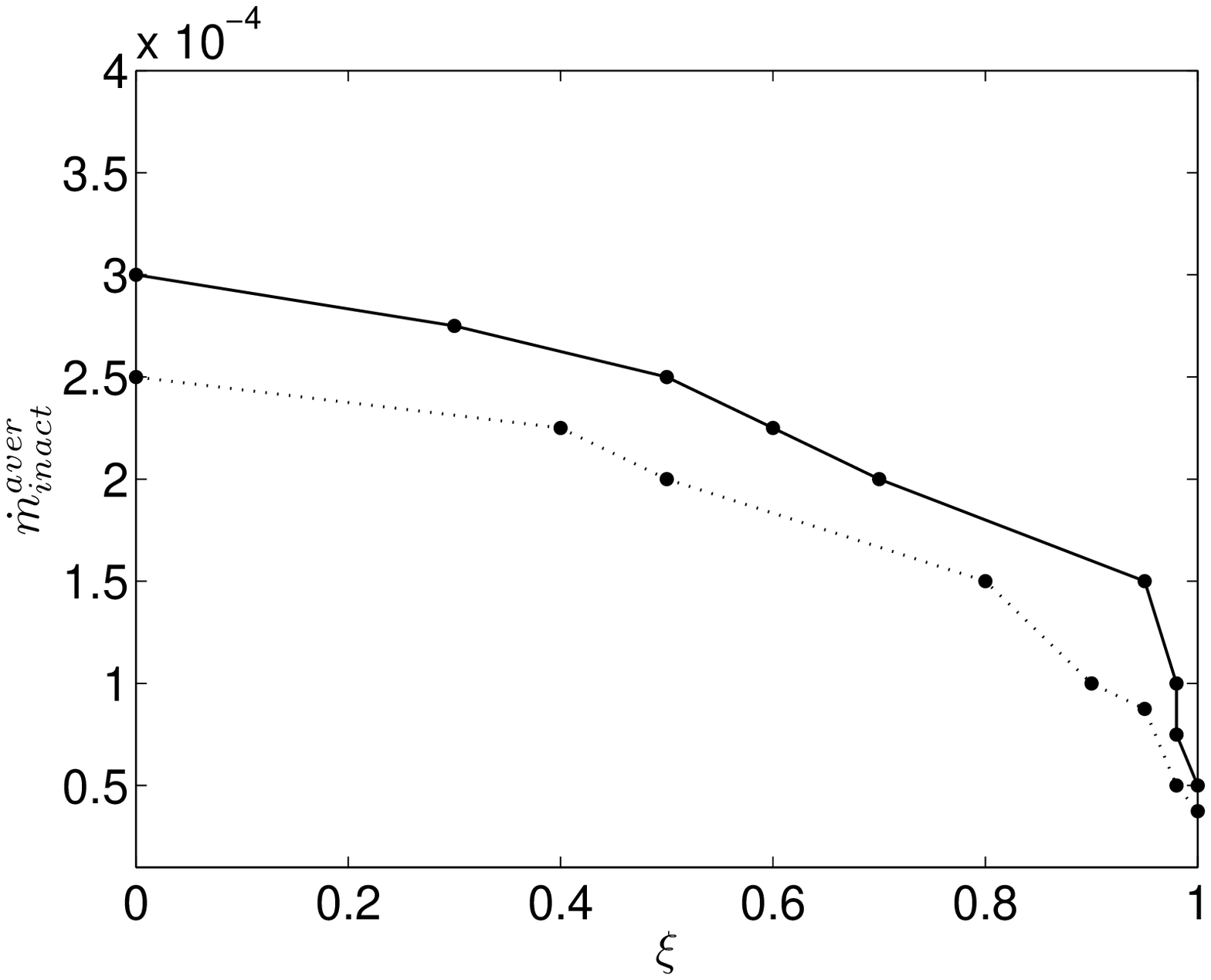}}
\figcaption{The constraints (upper limits) on the average mass accretion rates 
$\dot{m}_{\rm inact}^{\rm aver}$ as functions of the number ratio $\xi$ of the 
Compton-thick AGNs with $\log N_{\rm H}=24-25$ to those with $\log N_{\rm H}=23-24$. The solid line represents the result for 
$\rho_{\rm bh}^{\rm local}=2.5\times 10^{5} {\rm M_\odot~Mpc^{-3}}$, while 
the dotted line is for $\rho_{\rm bh}^{\rm local}=4.0\times 10^{5} {\rm M_\odot~Mpc^{-3}}$. 
\label{fig13}}\centerline{}

\section{Discussion}

As described in \S 2, we derive the black hole mass 
densities for active/inactive galaxies as functions of redshift from the HXLF assuming that massive black 
holes grow through accretion. Using different measurements on the local black hole mass 
density, the radiative efficiencies $\epsilon$ derived are required to be in the range of 
$\sim 0.07-0.12$ ($\epsilon\sim 0.11-0.12$, for $\rho_{\rm bh}^{\rm local}=2.5\times 10^5 {\rm M_\odot~Mpc^{-3}}$,  
and $\epsilon\sim 0.07-0.08$, for $\rho_{\rm bh}^{\rm local}=4.0\times 10^5 {\rm M_\odot~Mpc^{-3}}$, 
see Table 1 for detailed results). If part of the local black hole mass density 
is accumulated through the way other than accretion, the derived radiative efficiencies should 
be even higher than the present values. 
If $\dot{m}_{\rm inact}^{\rm aver}\la 3\times 10^{-4}$, less than $\sim$ 5 per cent of 
local black hole mass density was accreted in inactive galaxies.  

The black hole mass density $\rho_{\rm bh}^{\rm inact}(z)$ for inactive galaxies is 
derived in this work without considering Compton-thick AGNs. 
The total black hole mass density $\rho_{\rm bh}(z)$ derived in this work has been 
affected little by Compton-thick AGNs at low redshifts, because it is derived 
to match the local black hole mass density at $z=0$. Our calculations may 
underestimate the black hole mass density $\rho_{\rm bh}^{\rm act}(z)$ in active galaxies. 
Thus, the derived black hole mass density $\rho_{\rm bh}^{\rm inact}(z)$ for inactive galaxies should be overestimated 
(see Eq. \ref{rhobhinact}). Fortunately, the derived $\rho_{\rm bh}^{\rm inact}(z)$ has 
been affected very little at low redshifts, because only a small fraction of massive 
black holes are active at $z\la 1$ (see Fig. \ref{fig2}). The contribution to the XRB 
from inactive galaxies is dominated by that from the sources at low redshifts, which means 
that our main conclusions drawn from the XRB synthesis calculations will not be altered, 
even if Compton-thick AGNs are included in black hole mass density calculations.  

The derived radiative efficiencies $\epsilon$ would be higher than 
the present values listed in Table 1, if Compton-thick AGNs are included. If the ratio of Compton-thick to Compton-thin AGNs is 
$x$, a rough estimate gives the radiative efficiencies should be $1+x$ times of the present values, 
because only about $1/(1+x)$ of the local black hole mass density was accumulated in Compton-thin 
AGNs and inactive galaxies. Thus, the present derived radiative efficiencies $\epsilon$ are only 
the lower limits. If $x=0.6$ is adopted \citep*[e.g.,][]{t06}, the radiative efficiencies are about 1.6 times of the present 
values, i.e., $\epsilon\sim0.17$ 
for $\rho_{\rm bh}^{\rm local}=2.5\times 10^5 {\rm M_\odot~Mpc^{-3}}$, or $\sim 0.11$ 
for $\rho_{\rm bh}^{\rm local}=4.0\times 10^5 {\rm M_\odot~Mpc^{-3}}$, which implies that 
most massive black holes are spinning rapidly. \citet{erz02} 
suggested that a high average radiative efficiency $\epsilon\ga 0.15$ is required
from the XRB, and they concluded that most massive black holes must be spinning rapidly. 
Recently, \citet{w06} have estimated the average radiative efficiencies of a large sample of 
quasars selected from the Sloan Digital Sky Survey (SDSS), by combining their luminosity 
function and their black hole mass function, and found that quasars 
have average radiative efficiencies of $0.3-0.35$, which implies that 
the black holes are spinning very rapidly. Our present results seem to be consistent 
with those derived by \citet{erz02}, if suitable number of Compton-thick AGNs is considered.

\citet{md04} estimated that the average accretion rate
$\dot{m}^{\rm aver}_{\rm act}$ varies from 0.1 at $z\sim0.2$ to 0.4 at
$z\sim2$ from a large sample of SDSS quasars. 
The average accretion rate $\dot{m}_{\rm act}^{\rm aver}=0.1$ is adopted in this 
work to estimate the inactive galaxy black hole mass density $\rho_{\rm bh}^{\rm inact}(z)$.
We find that the value of $\dot{m}_{\rm act}^{\rm aver}$
may affect the resulting $\rho_{\rm bh}^{\rm inact}(z)$ very slightly at low redshifts, because
$\rho_{\rm bh}^{\rm inact}(z)\gg\rho_{\rm bh}^{\rm act}(z)$ except at high redshifts ($z\ga 2$, 
see Fig. \ref{fig2}). The derived black hole mass density $\rho_{\rm bh}^{\rm inact}(z)$ 
for inactive galaxies is mostly governed by the value of local black hole mass density 
$\rho_{\rm bh}^{\rm local}$ at low redshifts. The XRB is mainly contributed by the sources at 
low redshifts, which implies that the value $\dot{m}_{\rm act}^{\rm aver}$ only 
affect our calculations  of the XRB very little. 

A small fraction of the sources described by the HXLF with very high hard X-ray luminosities  
$\sim 10^{47-48}$ ergs~s$^{-1}$ are probably blazars \citep[e.g.,][]{u03}. For those blazars, 
the beamed X-ray emission from their jets may dominate over that from the accretion disks, 
which may lead to over-estimate of the black hole mass density accumulated through accretion. 
We perform the calculations of the black hole mass densities only for the sources with 
hard X-ray luminosities in the range of $10^{41.5}-10^{46}$ ergs~s$^{-1}$, and find that the 
fraction of total black hole mass density contributed by these X-ray luminous sources with 
$L_{\rm X}^{2-10~{\rm keV}}\ge 10^{46}$ ergs~s$^{-1}$ can be neglected ($\la 1\%$).  

We calculate the spectra of the RIAFs surrounding Schwarzschild black holes ($a=0$), and the Kerr black holes 
with $a=0.9$ using reasonable parameters for the RIAFs (see discussion in \S 3), respectively. 
For the RIAFs surrounding Schwarzschild black holes, we find that the X-ray spectra are 
dominated by the bremsstrahlung emission if the accretion rates are low (see Fig. \ref{fig5}),  
while the Comptonization is important for the RIAFs surrounding Kerr black holes (see Fig. 
\ref{fig6}). For a Kerr black hole, the accretion flow extends to a smaller radius and 
the gas density is higher at its inner edge compared with that for a Schwarzschild 
black hole, if they are accreting at the same rate. Thus, the Comptonization becomes 
significant for the RIAFs surrounding the Kerr 
black holes. Our calculations show that the peak energy of the spectrum decreases with increasing 
the accretion rate $\dot{m}_{\rm inact}^{\rm aver}$.  For the 
accretion flow with a higher $\dot{m}_{\rm inact}^{\rm aver}$, the density at its inner edge is 
also higher that leads to a lower turnover frequency for the synchrotron radiation from the 
accretion flow, and then leads to a lower energy peak produced by the Comptonization, because the electron 
temperatures of the accretion flows at their inner edges are roughly similar for different values 
of $\dot{m}_{\rm inact}^{\rm aver}$.

To compare our calculations of the XRB contributed by active/inactive galaxies 
with the observed XRB, we find that the average accretion rate of inactive 
galaxies $\dot{m}_{\rm inact}^{\rm aver}$ should be less than $0.001$, otherwise, 
the calculated XRB will surpass the observed XRB. Our calculations show that 
$\dot{m}_{\rm inact}^{\rm aver}\la 2-3\times 10^{-4}$ is required by the observed 
XRB (see Figs. \ref{fig9} and \ref{fig10}).  It is interesting to find that the spectral shape of 
the observed XRB (peaked at $\sim 30$ keV) cannot be well fitted by using any value of 
$\dot{m}_{\rm inact}^{\rm aver}$ for Schwarzschild 
black holes, while both the observed intensity and the spectral shape of the XRB can be 
well reproduced if $\dot{m}_{\rm inact}^{\rm aver}\simeq 2.5\times 10^{-4}$ is adopted for the 
Kerr black holes with $a=0.9$ (see Fig. \ref{fig10} for 
$\rho_{\rm bh}^{\rm local}=4.0\times 10^{5} {\rm M_\odot~Mpc^{-3}}$). 
This implies that most massive black holes are spinning rapidly, which seems to be consistent 
with the relatively high average radiative efficiencies derived from the XLF 
(see discussion in the second paragraph of this section). 
A slightly higher average accretion rate $\dot{m}_{\rm inact}^{\rm aver}\simeq 3.0\times 10^{-4}$ is 
required for the case of $\rho_{\rm bh}^{\rm local}=2.5\times 10^{5} {\rm M_\odot~Mpc^{-3}}$. 
Note that the best-fitted value $\dot{m}_{\rm inact}^{\rm aver}$ does not 
depend linearly on  $1/\rho_{\rm bh}^{\rm local}$, because the luminosity has a non-linear 
dependence of accretion rate for RIAFs \citep[e.g.,][]{wc06}.  Here, we have neglected the 
contribution of the Compton-thick AGNs, as it is still unclear how many Compton-thick AGNs are 
in the universe \citep[e.g.,][]{u03}. Thus, these results should be only regarded as 
the upper limits on accretion rates in inactive galaxies. We further use a parameter 
$\xi$ to describe the number ratio of 
Compton-thick AGNs with $\log N_{\rm H}=24-25$ to those with $\log N_{\rm H}=23-24$.  
For the case of $\xi=0.5$, our XRB synthesis calculation shows that the XRB 
contributed by active and inactive galaxies (type I/II Compton-thin AGNs
$+$Compton-thick AGNs$+$RIAFs in inactive galaxies) can match the observed XRB 
for mass accretion rates $\dot{m}_{\rm inact}^{\rm aver}\simeq 2.0\times 10^{-4}$  
(see Fig. \ref{fig12} for the case of $\rho_{\rm bh}^{\rm local}=4.0\times 10^{5} {\rm M_\odot~Mpc^{-3}}$). 
A slightly higher accretion rate $2.5\times 10^{-4}$ is required for the case of 
$\rho_{\rm bh}^{\rm local}=2.5\times 10^{5} {\rm M_\odot~Mpc^{-3}}$. If the contribution from 
the RIAFs in inactive galaxies is not considered, the same number of the Compton-thick AGNs 
with $\log N_{\rm H}=24-25$ as those with $\log N_{\rm H}=23-24$ is required to reproduce 
the observed XRB (see the yellow line in Fig. \ref{fig12}).  
In Fig. \ref{fig13}, we find that the contribution 
to the XRB from inactive galaxies can be neglected compared with that from Compton-thick 
AGNs, if $\dot{m}_{\rm inact}^{\rm aver}\la 10^{-4}$. Corresponding to such constraints on the 
average accretion rates for inactive galaxies, we find that less than $\sim$5 per cent, or even 
less than $\sim$2 
per cent if Compton-thick AGNs are considered, of 
the local massive black hole mass density was accreted during the radiatively inefficient 
accretion phases (see Table 1). The constraints of the XRB can only give upper limits on 
$\dot{m}_{\rm inact}^{\rm aver}$ for inactive galaxies, and we cannot rule out the possibility 
that inactive galaxies are accreting at rates significantly lower than $10^{-4}$. This may be 
resolved by deep surveys on faint X-ray sources in hard X-ray bands, which is beyond the 
scope of this paper.

The detailed ADAF spectral calculations by \citet{d99} showed that many sources at
redshift $z\sim 2-3$ with ADAFs accreting at the rates close to
the critical value would be required to reproduce the observed spectral shape 
of the XRB with an energy peak at $\sim 30$~keV. However, \citet{cao05}'s spectral calculations 
for RIAFs accreting at the critical value showed higher energy peaks at several hundred keV, 
which are unable to reproduce the observed spectral shape of the XRB (see Fig. 2 in Cao, 2005). The
reason is that \citet{cao05} adopted a larger $\delta$ than that used by \citet{d99} for 
traditional ADAFs, which leads to higher electron temperatures of the accretion flows 
\citep[see][for details]{cao05}. 
\citet{d99}'s calculations are limited to the spectral shape, which have not included the  
contribution of the ADAFs in all galaxies to the XRB quantitatively, and the contribution 
of normal bright AGNs has not been considered either.  Our present 
calculations show that the spectra of the RIAFs surrounding Kerr black holes have peaks 
at around 10-100 keV, which depend sensitively on accretion rates (see Fig. 4). We find that 
our model calculations can fit both the intensity and the spectral shape of the XRB very well, 
if the RIAFs surrounding the Kerr black holes with $a\sim 0.9$ in inactive 
galaxies are accreting at $\dot{m}_{\rm inact}^{\rm aver}\simeq 1-3\times 10^{-4}$.  
\citet{cao05} calculated the contribution to the XRB from the RIAFs accreting 
at the critical rate, and found that the timescale for the sources accreting at the 
critical rate should be much shorter than the bright AGN lifetime. In that work, it was found 
that the spectral shape of the XRB cannot be reproduced even the RIAFs' contribution is included. 
Our present calculations imply the contribution from the RIAFs in the inactive galaxies 
accreting at $\simeq 1-3\times 10^{-4}$ should dominate over that from those accreting 
at the critical value. It means that the timescale of AGNs accreting at around the critical value 
should be even shorter than that given by \citet{cao05}.

In this work, we have to assume the black holes in all inactive galaxies are accreting at an average 
rate $\dot{m}_{\rm inact}^{\rm aver}$, because we are almost ignorant of how the accretion rate 
evolves with time in active/inactive galaxies. A more realistic possibility may be that the 
massive black holes in inactive galaxies are accreting at different rates from far below 
$\sim 10^{-5}$ up to the critical value of $\sim 0.01$, of which the stacked spectra (combined  
with the contribution from the type I/II AGNs) may reproduce the observed XRB. The spectrum of the 
RIAF accreting at the critical value is peaked at $\sim 100$~keV, due to the saturated 
Comptonization \citep[see Fig. 1 in][]{cao05}. These RIAFs are X-ray luminous, and therefore 
much more inactive galaxies accreting at very low rates ($\ll\dot{m}_{\rm inact}^{\rm aver}$) are required 
to reproduce the intensity of the XRB. However, the spectral shape of the XRB with an energy peak 
$\sim 30$~keV cannot be fitted in this way, because the energy peaks of the RIAF spectra 
are around several hundred keV while they are accreting at $\dot{m}\la 10^{-4}$ 
(see Figs. \ref{fig5} and \ref{fig6}). It means the contribution of the RIAFs accreting at 
around the critical value to the XRB should be unimportant compared with other 
inactive galaxies, which, again, suggests that the accretion rates of 
the RIAFs in inactive galaxies should decreases very rapidly after the accretion mode transition. 
It provides evidence of the feedback on the circumnuclear gases from the nuclear 
radiation, which is qualitatively  consistent with the numerical simulations \citep*[see Fig. 2 in][]{d05}.

The RIAF may have winds, and a power-law radius-dependent accretion rate, 
$\dot{m}(r)\propto r^{-p_{\rm w}}$ ($p_{\rm w}>0$), is assumed, 
though the detailed physics is still unclear \citep{bb99}. As most X-ray emission is from the
inner region of the accretion flow near the black hole, the X-ray spectrum of a RIAF is similar
to a RIAF with winds, provided they have the same accretion rate at their inner edge. In this
work, we are investigating the growth of black holes, so we focus on the accretion rate at 
the inner edge of the accretion flow. Thus, the conclusions will not be altered, even if 
winds are present in the RIAFs.

\acknowledgments  This work is supported by the National Science Fund for Distinguished
Young Scholars (grant 10325314), and the NSFC (grant 10333020).

\begin{deluxetable}{cccccccc}
\tabletypesize{\scriptsize} \tablecaption{The summary of different models}
\tablewidth{0pt} \tablehead{ \colhead{Model} &
\colhead{$\dot{m}_{\rm inact}^{\rm aver}$} & \colhead{$\epsilon$}
& \colhead{${\rho_{\rm bh}^{\rm local}}^{\rm a}$} & \colhead{$\rho_{\rm bh}^{\rm B}(0)/\rho_{\rm bh}^{\rm local}$}
& \colhead{$\epsilon$}
& \colhead{${\rho_{\rm bh}^{\rm local}}^{\rm a}$} & \colhead{$\rho_{\rm bh}^{\rm B}(0)/\rho_{\rm bh}^{\rm local}$} }
\startdata
  A & $1.0\times 10^{-3}$ & 0.122 & $2.5\times10^{5}$ & 0.162 & 0.080 & $4.0\times10^{5}$ & 0.162  \\
  B & $3.0\times 10^{-4}$ & 0.110 & $2.5\times10^{5}$ & 0.051 & 0.071 & $4.0\times10^{5}$ & 0.051 \\
  C & $2.0\times 10^{-4}$ & 0.108 & $2.5\times10^{5}$ & 0.034 & 0.070 & $4.0\times10^{5}$ & 0.035  \\
  D & $1.0\times 10^{-4}$ & 0.106 & $2.5\times10^{5}$ & 0.017 & 0.069 & $4.0\times10^{5}$ & 0.017  \\
  E & $0.0$ & 0.105 & $2.5\times10^{5}$ & 0.0 & 0.068 & $4.0\times10^{5}$ & 0.0  \\ 
\enddata
 \tablenotetext{a}{in units of $\rm M_\odot~Mpc^{-3}$.}
\end{deluxetable}


\end{document}